\documentclass{article}

\usepackage{amssymb}

\usepackage{epsfig}

\usepackage{lscape}

\usepackage{graphicx,psfrag,amsmath}

\usepackage{yfonts}

\begin{document}

\def\beq#1\eeq{\begin{equation}#1\end{equation}}
\def\beql#1#2\eeql{\begin{equation}\label{#1}#2\end{equation}}

\def\bea#1\eea{\begin{eqnarray}#1\end{eqnarray}}
\def\beal#1#2\eeal{\begin{eqnarray}\label{#1}#2\end{eqnarray}}

\newcommand{\Z}{{\mathbb Z}}
\newcommand{\N}{{\mathbb N}}
\newcommand{\C}{{\mathbb C}}
\newcommand{\Cs}{{\mathbb C}^{*}}
\newcommand{\R}{{\mathbb R}}
\newcommand{\intT}{\int_{[-\pi,\pi]^2}dt_1dt_2}
\newcommand{\cC}{{\mathcal C}}
\newcommand{\cI}{{\mathcal I}}
\newcommand{\cN}{{\mathcal N}}
\newcommand{\cE}{{\mathcal E}}
\newcommand{\cA}{{\mathcal A}}
\newcommand{\xdT}{\dot{{\bf x}}^T}
\newcommand{\bDe}{{\bf \Delta}}

\def\ket#1{\left| #1\right\rangle }
\def\bra#1{\left\langle #1\right| }
\def\braket#1#2{\left\langle #1\vphantom{#2}
  \right. \kern-2.5pt\left| #2\vphantom{#1}\right\rangle }
\newcommand{\gme}[3]{\bra{#1}#3\ket{#2}}
\newcommand{\ome}[2]{\gme{#1}{#2}{\mathcal{O}}}
\newcommand{\spr}[2]{\braket{#1}{#2}}
\newcommand{\eq}[1]{Eq\,\ref{#1}}
\newcommand{\xp}[1]{e^{#1}}

\newcommand{\tend}[1]{$10^{#1}$}
\newcommand{\tennd}[1]{10^{#1}}
\newcommand{\nrd}[2]{${#1}\times{10^{#2}}$}
\newcommand{\nrnd}[2]{{#1}\times{10^{#2}}}

\def\limfunc#1{\mathop{\rm #1}}
\def\Tr{\limfunc{Tr}}

\def\dr{detector }
\def\drs{detectors }
\def\drsn{detectors}
\def\drn{detector}
\def\dtn{detection }
\def\dtnn{detection}

\def\pho{photon }
\def\phon{photon}
\def\phos{photons }
\def\phosn{photons}
\def\mmt{measurement }
\def\an{amplitude}
\def\a{amplitude }
\def\co{coherence }
\def\con{coherence}

\def\st{state }
\def\stn{state}
\def\sts{states }
\def\stsn{states}

\def\cow{collapse of the wavefunction }
\def\de{decoherence }
\def\den{decoherence}
\def\dm{density matrix }
\def\dmn{density matrix}

\newcommand{\mop}{\cal O }
\newcommand{\dt}{{d\over dt}}
\def\qm{quantum mechanics }
\def\qms{quantum mechanics }
\def\qml{quantum mechanical }

\def\qmn{quantum mechanics}
\def\mmtn{measurement}
\def\pow{preparation of the wavefunction }

\def\me{ L.~Stodolsky }
\def\T{temperature }
\def\Tn{temperature}
\def\t{time }
\def\tn{time}
\def\wfs{wavefunctions }
\def\wf{wavefunction }
\def\wfn{wavefunction} 
\def\wfsn{wavefunctions}
\def\wvp{wavepacket }
\def\pa{probability amplitude } 
\def\sy{system } 
\def\sys{systems }
\def\syn{system} 
\def\sysn{systems} 
\def\ha{hamiltonian }
\def\han{hamiltonian}
\def\rh{$\rho$ }
\def\rhn{$\rho$}
\def\op{$\cal O$ }
\def\opn{$\cal O$}
\def\yy{energy }
\def\yyn{energy}
\def\yys{energies }
\def\yysn{energies}
\def\pz{$\bf P$ }
\def\pzn{$\bf P$}
\def\pl{particle }
\def\pls{particles }
\def\pln{particle}
\def\plsn{particles}

\def\plz{polarization  }
\def\plzs{polarizations }
\def\plzn{polarization}
\def\plzsn{polarizations}

\def\sctg{scattering }
\def\sctgn{scattering}
\def\sctgs{scatterings }
\def\sctgsn{scatterings}

\def\prob{probability }
\def\probn{probability}

\def\om{\omega} 

\def\hf{\tfrac{1}{2}}
\def\hft{\tiny \frac{1}{2}}

\def\zz{neutrino }
\def\zzn{neutrino}
\def\zzs{neutrinos }
\def\zzsn{neutrinos}

\def\zn{neutron }
\def\znn{neutron}
\def\zns{neutrons }
\def\znsn{neutrons}

\def\hf{\tfrac{1}{2}}

\def\csss{cross section }
\def\csssn{cross section}
\def\xrn{x-ray nucleide }
\def\xrnn{x-ray nucleide}
\def\xr{x-ray }
\def\xrs{x-rays }
\def\wv{wavelength }
\def\wvn{wavelength}

\def\intf{interferometry }
\def\intfn{interferometry}

\def\ran{radionuclide }
\def\rann{radionuclide}
\def\rans{radionuclides }

\def\dkm{dark matter }
\def\dkmn{dark matter}

\def\gvtl{gravitational }
\def\gvtln{gravitational}

\def\bh{black hole }
\def\bhn{black hole}
\def\bhs{black holes }
\def\bhsn{black holes}

\def\dph{$\delta \phi$ }
\def\dphn{$\delta \phi$}
\def\dphg{$\delta \phi_G$}
\def\dphgn{$\delta \phi_G$}

\def\gr{general relativity }
\def\grn{general relativity}

\def\mua{$\mu arcsec$ }

\def\los{line-of-sight }
\def\losn{line-of-sight}

\title{
 Observational Aspect of Black Hole Dark Matter}

\author{ 
L. Stodolsky,\\
Max-Planck-Institut f\"ur Physik
(Werner-Heisenberg-Institut)\\
F\"ohringer Ring 6, 80805 M\"unchen, Germany}

\maketitle

\begin{abstract}
 
 Advances in high angular resolution astronomy
  make it conceivable  that \bh \dkm could be detected via angular
 deviation effects.

Assuming the \dkm in the galaxy is made of solar mass \bhsn, 
 there is a non-trivial probability that
 a \los through the galaxy,
leads to \mua's deviations,  a value that has been
 discussed for various astronomical projects.

In cosmology the effects are magnified by an increased
density at early times and an
opening of angles due to redshift. We stress the interest of the
resolution in observations on the CMB and emphasize the distinction
between this and  the power spectrum.
 If the \dkm is made of primordial \bhsn,
  present at  the CMB,
 random deflections of the CMB \phos lead to
 a limit on the angular  resolution, approximately
$\nrnd{3}{-7} \sqrt{M/M_\odot}\, rad$, with $M$ the mass of
the \bhsn.
 Using the  resolutions of
$\sim\tennd{-3} rad$ demonstrated in  observations of the
``acoustic
peaks '' then implies the limit
 $(M/M_\odot)\lesssim \tennd{7}$. While this large value seems
uninteresting,  improved resolutions would lead
to significant exclusion limits or conceivably the discovery of 
primordial
\bhsn.
 
\end{abstract}

\section{Introduction}
The problem of the nature of the \dkm is and remains one of the
primary and most fascinating questions of contemporary science.
While the search for elementary \pl \dkmn, particularly  
by means of the cryogenic technique\cite{cryo}, seems at present
the most plausible path to its elucidation, some other
possibilities have been discussed.

One of the most interesting of these possibilities is that the \dkm
is made of \bhsn,  primordial objects presumably  originating from
the very early universe. These have not been
definitively excluded and it has been suggested that they
could help in the
understanding of structure formation in cosmology\cite{carr}.

 Discussions concerning this possibility have necessarily
relied on indirect arguments. But it certainly would be more
satisfying if
there were direct observational evidence, for or against the \bh
hypothesis. Here we would like to consider how it may be 
 possible to obtain such evidence by means of high angular
resolution observations.

For a preliminary orientation we first consider the situation for
the galaxy, and then turn to the CMB (Cosmic Microwave Background
),
 where general relativistic
effects lead to large enhancements.

\section{Deflection  and Distance} \label{dd}

Our arguments are based on the deflection of light by massive
bodies, as in the famous bending of light by the sun. If \bhs make
up the \dkmn, the interstellar or intergalactic 
medium would be very ``lumpy'' on short distance scales, as opposed
to \dkm made of elementary \plsn. 

A standard result in General Relativity \cite{answer}  is that
 the deflection angle $\delta \alpha$  of a  ray
passing a gravitating
object of mass M at impact parameter $b$, is
\beql{dfl}
\delta \alpha=\frac{4GM}{c^2b} =2\frac{r_s}{b}=\frac {6.0 \, km
}{b}\, ( M/M_\odot)
=\frac {\nrnd{6.3}{-13} \, ly }{b}\, ( M/M_\odot) \,,
\eeql
where we express $M$
 in terms of the Schwarzschild radius $r_s=2GM$ and normalize to
the mass of the sun, $M_\odot$, where one has   $r_s=3.0\, km$.
Here and in the following we use $c=1$ units.

Thus for  a deflection of $\delta \alpha$ or
more  there is  a certain impact paramter $b$, to which we can
associate a `` \csss''  
\beal{css}
\pi b^2=\pi  \biggl(\frac{2 r_s}{\delta \alpha
}\biggr)^2&=&\nrnd{1.1}{2}
\biggl(\frac{1}{\delta\alpha}\biggr)^2 (M/M_\odot)^2
km^2~~~~~~~~~~~~~~~~~~~~~~~~\\
\nonumber
&=&\nrnd{1.3}{-24}
\biggl(\frac{1}{\delta
\alpha}\biggr)^2 (M/M_\odot)^2 ly^2\,.
\eeal
This is the \csss for having, in a single passage near a massive
object, a deflection $\delta \alpha$ or more. 

(Here and in the following we speak of the \bhs as having a single,
unique mass; if  instead there is a spectrum of masses, $M$ should
be understood as the average mass. )
   
 If we consider $\delta \alpha$'s on the order of a \mua $=
\nrnd{4.8}{-12} rad$,
as in long baseline interferometry \cite{lbi}, this \csss can be
$\sim ly^2$ for a solar mass object. Equivalently, passages at
distances $\sim ly$ are potentially observable given such
resolutions.

\section{Probability of a  $\delta \alpha$}\label{pda}
One may turn \eq{css} into   the probability for
a deflection $\delta \alpha$ (or more) for a \pho traveling a
certain path   by multiplying by the column density
\beql{prob1}
Prob_{\delta \alpha}=cross\, section\, \times column\, density=\pi
\bigl(\frac{2r_s}{\delta
\alpha}\bigr)^2\times \rho_2\,, 
\eeql
where the column density  $\rho_2$  is 
 the two-dimensional  density of the total number
of \bhs along the line-of-flight, projected on the
 perpendicular plane. (We are taking 
 $Prob_{\delta \alpha}$ to be small, a value approaching one
 implies multiple encounters. )

The $\rho_2$  can in turn be expressed in terms of the presumably
approximately known \dkm mass column density and the unknown
\bh mass $M$ as 
\beql{2d}
\rho_2=\mu/M\,,
\eeql
where the \dkm mass column density is expressed as $\mu/unit~area$,
with $\mu$  a mass.

Since $r_s^2\sim M^2$ one sees that \eq{prob1} is proportional to
$ M$ and vanishes  as $M\to 0$.
 We are working with a  constant, given,  \dkm mass density, so in
the limit
$M\to 0$ the medium becomes perfectly smooth. This is in accord
with the expectation that 
the deflection effect depends on how   
  ``lumpy'' the medium is and would be  
  expected to vanish, say, when the \dkm  consists of
a smooth ``gas'' of elementary \pls.

\section{Galaxy}

We first apply the above estimates to the case of the galaxy.
In his original discussion of `Machos'(Massive compact halo
objects),
  Paczynski \cite{machos} 
briefly mentioned this possibility of the 
direct observation of angular deviations,  but dismissed it  as
immeasurably small.
However,
since that time there have been great improvements in angular
resolutions; also
  a brief discussion for the galaxy  will allow us
  to establish the ideas without the complications of
\grn.

While an accurate calculation of $\rho_2$ requires an integration
of the \dkm density along the
flight path, we can crudely estimate it for a source in
the Milky Way (or for rays traveling through a nearby galaxy ) by
taking the presumed \dkm density near the earth  of
$0.4\, GeV/cm^3= \nrnd{7.1}{-28}kg /cm^3=\nrnd{6.0}{26}kg
/ly^3$\cite{pdg} and a typical
galactic travel
distance of \tend{4} light years.
One thus obtains
\beql{2da}
\rho_2\approx \frac{1}{M}\bigl(\nrnd{6.0}{26}kg/ly^3\bigr) \times
\tennd{4} ly = 3.0\times (M_\odot/M)/ly^2                    
~~~~~~~~galaxy
\eeql

Combining \eq{2da} and \eq{prob1}, one has the probability of
a deflection of $\delta \alpha$  or more in
a typical passage through the galaxy 
\beql{prob4}
Prob_{\delta \alpha}\approx\nrnd{3.9}{-24}(1/\delta
\alpha)^2(M/M_\odot) =  \nrnd{1.6}{-1}\times (1/\delta
\alpha')^2
(M/M_\odot)~~~~~~~~~~~~galaxy\,.
\eeql
In the first writing $\delta \alpha$ is in radians, while in the
second $\delta \alpha'$ is in \mua.
Thus if the \dkm is made up of solar mass (or more) \bhsn,
 a ray crossing the galaxy has a
substantial chance of a  several \mua deflection.
 This estimate is of course
quite approximate, and will depend  on whether the
flight path is through regions of high or low \dkm density.

The next question  is how this very small angular deflection might
be detected, given that there are generally larger effects from the
motion of the source and the earth.
To obtain an observable signal, one might look for
a motion  of $\delta \alpha$.  The \bhn,
like other
objects in the galaxy, will be moving with some velocity. This
velocity will lead to a
change in $b$ in \eq{dfl} and so a change in
$\delta \alpha$. Since we are considering passages at distances
$b$ on the order of a $ly$ or less and the typical 
velocity of objects  in the galaxy 
$v\approx \nrnd{2}{-3}$ corresponds to a distance traveled $\approx
\nrnd{2}{-3} ly$ in one year, it is possible that a significant
change in $b$ can occur over a timescale 
of years. The
  presence of the \bhs  would  be thus
signaled
as a small `noise', with a timescale
 on the order of  years,  on the more smoothly varying angular
positions of
the  sources.

 A shorter timescale for the `noise' could
be obtained by raising the threshold of \dtn for $\delta \alpha$,
  implying smaller
$b$, but then a reduced probability $Prob_{\delta \alpha}.$ This
behavior
with respect to threshold would be useful in establishing the
reality of
a possible posit;ive signal.

\section{CMB As  Source}

We now turn to the CMB, the most remote soure imaginable.
This turns out to be of fundamental interest since, as we shall see,
it is in principle possible, although observationally difficult, to set 
significant limits, or perhaps to discover primrodial \bhs through 
observations on the CMB.

While there have been direct observations concerning \bhsn,
 these observations  have been
 in the nearby universe, as in \cite{lbi}, or through the observation 
of accretion disks, or  
 by rapid motions of stars in the center of the Milky Way \cite{nobel}.  On the other
hand, arguments concerning {\it primordial} \bhs have mainly revolved  around
their effects on cosmic evolution as in \cite{carr}. In contrast,
  our method
would amount to a direct observation.

Very briefly our point is, as familiar in  particle or atomic  physics,
that  \sctg centers between a source of radiation and a detector or 
 observer lead
to a loss of angular resolution, We apply this point to the
presence of \bhs between us and the CMB. 
 Thus instead of considering
a particular object as the source, we examine
 the typical deflection of a \pho from the CMB
and argue that, due to the randomness of this deflection,  smaller
angular features will be   washed out. That is, the presence of
primordial
 \bhs implies a limit on the possible angular resolution in
 CMB observations.  

An important aspect of the calculation, as seen below, is that the 
effect largely  arisess from deflections at early times, when the
cosmological scale factor $a(t)$ is small. Without this enhancement the 
effect would be undetectably small. Hence a positive observation would
strongly suggest that the deflecting objects are primordial. 

 This point concerning observations on the CMB distinguishes
primordial \bhs from 
other  compact gravitating object that might form in the course
of cosmic evolution, but not present at the CMB,  such as ``Machos'', ``Brown
Dwarfs'', and so forth.
With the present
understanding of
cosmology,
such compact objects would exist after
 the formation of structure, and  would not
be present at the CMB. On the
other hand, primordial  \bhs  would presumably arise from 
 very early times \cite{carr}. 

The perhaps surprising conclusion  that angular deviations
might be observable results from some
 general relativistic effects in cosmology, namely
 an  opening of the  deflection angle seen by the observer,
and  the high density  of \dkm  at earlier times. These  lead to a
large quantitative enhancement  relative to \eq{prob4}. As said,
 the enhancement is large and  occurs at early times or high
redshift and  distinguishes the primordial \bhs effect
from  those originating 
  from density concentrations developing out of a smooth background.
 The later will necessarily lead
to much smaller effects since there is not the enhancement at early
times.

\subsection{Opening of Angles}
A deflection angle produced at   high redshift is
 magnified when observed ``now'', due to the cosmological redshift.
  We work in the simplest  FRW model
 for cosmology, where one has the
scale factor $a(t)=(t/t_{now})^{2/3}$ after the formation of the
CMB, with  $t_{now}=\nrnd{2.9}{17} s$.  (Note this notation introduces
a factor $3/2$ between $t_{now}$ and the hubble parameter. )

 In the local frame of the \bhn, the \pho before
deflection has
the 4-vector $k$ and after deflection $k'$. The scalar 4-product is
$kk'=\omega^2(1-cos\delta\alpha)\approx \hf \omega^2 (\delta
\alpha)^2$. We take the two \phos to have the same frequency and to
differ by a small angle, as for  \phos deflected
by a  massive \bhn. 
 A scalar product is conserved under  parallel transport, and since
the
 \phos come to us by free fall or  parallel transport,
a deflection
$\delta \alpha_t$ at cosmological time $t$ has the relation to the 
$\delta \alpha$ at the time
 `now', $(\omega_t \delta \alpha_t)^2=(\omega_{now} \delta
\alpha)^2 $.

Therefore the deviation  angles are in the ratio of the frequencies
and so are
increased by the redshift. 
\beql{plla}
 \delta \alpha= \frac{1}{a(t)}\times \delta \alpha_{t}  \,
\eeql
where $a(t)$ is the scale factor at the cosmological time $t$ of
the \sctgn.

This means that a deflection $\delta \alpha$  `now' originates from
a smaller deflection at high redshift, and since these will occur
at larger
impact parameter, they have a higher probability than
would be the case without this angular effect.
(This logic can also be used to show that a nonrelativistic
transverse motion has only a small
effect on $\delta \alpha$; also applicable for the galaxy case.)

\subsection{Increase of Density}
A second effect is the increased density of the \dkm at early
times, which
is  expected to  vary as $1/a^3$, such that number density at time
$t$
is   $\rho /a^3$,  where $\rho$ is the present number
density. 
   The contribution to the  column density
  from a cosmic time interval
$dt$ is then

\beql{drh}
  d\rho_2= 
\rho  \frac{1}{a^3} dt  = \frac{\mu}{M} \frac{1}{a^3} dt \,,
\eeql
where $\mu$ is the cosmological \dkm mass density at present
\cite{zack}.

\subsection{Probability Integral}
Taking these two effects into account 
\beql{nearu}
d Prob_{\delta  \alpha}=\pi b^2(t)  d\rho_2
= \pi  \bigl(\frac{2 r_s}{(a(t)\delta \alpha}\bigr)^2  d\rho_2\,,
\eeql
using   \eq{plla} and \eq{dfl} to give   $b(t)= 2 r_s/(a(t)\delta
\alpha) $. 
Combining with \eq{drh}:
\beql{nearu1}
\int d Prob_{\delta  \alpha}
=  \frac{\mu}{M} \pi  \bigl(\frac{2 r_s}{\delta \alpha}\bigr)^2\int
\frac{1}{a^5}
dt  \,,
\eeql

The factor $1/a^5$ will lead to large enhancement over the simple
dimensional factors:
\beql{aint}
\int^{now}_{cmb}dt \frac{1}{a^5}=t_{now}  \frac{3}{7}(x^{7/3}-1)
= \nrnd{7.2}{9}\times t_{now}
\eeql 
with $x=t_{now}/t_{cmb}=\nrnd{2.4}{4}$,
integrating over the time since
$t_{cmb}=\nrnd{3.8}{5}y$.

We are left with the task of evaluating the dimensional factor
\beql{csts}
\frac{\mu}{M} \pi  \bigl(\frac{2 r_s}{\delta \alpha}\bigr)^2
t_{now}
\eeql
The ``\csss '' was evaluated in \eq{css} 
\beql{xsec}
\pi  \bigl(\frac{2 r_s}{\delta \alpha }\bigr)^2=
\nrnd{1.3}{-24}
\biggl(\frac{1}{\delta
\alpha}\biggr)^2 (M/M_\odot)^2 ly^2
\eeql
The remaining factor $ \frac{\mu}{M}  t_{now}$
can be interpreted as the column density for a distance  $t_{now}$
($\sim$ Hubble distance) at the present \dkm density, without the 
general relativistic  effects. To evaluate this
 we take the present \dkm  mass density $\mu$ at $1/4$ the
critical value \cite{pdg}: 

\beql{lyden}
\mu= \nrnd{1.0}{-9}  M_\odot/ ly^3\,,
\eeql
 which together  with 
$t_{now}=\nrnd{9.2}{9} yr$
yields
\beql{pref}
\frac{\mu}{M}  t_{now}={9.2} \times
(M_\odot/M)/ly^2
\eeql

Interestingly,  this is about the same as the estimate for the
galaxy in
\eq{2da}; a factor \tend{6} in the
densities has been cancelled by a similar factor in the distances.

Thus the main difference vis-a-vis the galactic effects arises from
the large factor of \eq{aint},  from the $1/a^5$ behavior.
 The dominant effect thus comes from small $a$. The necessary
 ``clumpiness'', noted at the end of section \ref{pda}, originates
from
the primordial \bhs  present at early times.
 This is quite different from accretion or
gravitational accumulation mechanisms starting from an originally
smooth
medium and gradually building up at much later times. Such
late-time buildup
of density contrasts  would
not benefit greatly from the small $a$ magnification and should
lead to much
smaller results than those given here. The  primordial \bhsn,
 if they exist, would naturally also be expected to grow by
accretion, but,
as said, the dominant effect will come from their size  
at around the time of the formation of the CMB.

To finally estimate the integrated \eq{nearu1} for \phos from the
CMB we put together \eq{aint}, \eq{xsec}, and \eq{pref},  to find
\beal{tot}
Prob_{\delta \alpha}
=\nrnd{8.6}{-14} \biggl(\frac{1}{\delta \alpha }\biggr)^2 
(M/M_\odot)
=\nrnd{3.7}{9} \biggl(\frac{1}{\delta \alpha' }\biggr)^2 
(M/M_\odot)~~~~~CMB\,
\eeal
where again  $\delta \alpha $ is in radians while $\delta \alpha'$
is in \mua.

\subsection{Discussion}

To convert these results into a possibly observable effect,
consider the question of the angular resolution possible in
observations on the CMB.
When  $Prob_{\delta \alpha}$ is of order one, a \pho from a  point
on the CMB, will undergo a deflection ${\delta \alpha}$ with high
probability. 
Since these deflections  are random in direction, the angular
position of points will be altered  by ${\delta \alpha}$.
This implies a  limit on the angular resolution, set by the
requirement that $Prob_{\delta \alpha}$ be less than one. Setting
$Prob_{\delta \alpha}\approx 1$ in \eq{tot} gives
\beql{resl}
\delta \alpha_{lim} \approx
\nrnd{2.9}{-7}\sqrt{(M/M_\odot)}~~~~~~~~\delta \alpha'_{lim}
\approx \nrnd{6.1}{4}
\sqrt{(M/M_\odot)}   \,,
\eeql
as an approximate limit for the best obtainable resolution in 
observations on the CMB, in radians or \mua respectively.

To see how this might work in practice, we take the observation  
of  the ``acoustic
peaks'' in the  temperature fluctuations of the CMB. (We stress
that here we are not interested in the temperature fluctuations
themselves but simply in the demonstration of an angular
resolution.)
These features have been observed out to $l\sim 1000$,  implying 
angular resolutions of about  $\sim \tennd{-3}rad$. Using \eq{resl}
this implies an approximate upper limit on the mass of possible
\bhs 
\beql{mlim}
(M/M_\odot)\lesssim \tennd{7} 
\eeql
 The examples of very massive \bhs found up to the present time
are in the  range of millions of solar masses as in  \cite{lbi}
\cite{nobel}, and the existence of
 even larger objects should have been evident. For a discussion
of such limits see \cite{dolgov}.

Thus \eq{mlim} does not seem a very stringent limit,
 However, the argument does suggest
that higher resolution observations on the CMB could lead to
significant restrictions or perhaps even positive evidence for \dkm
\bhsn. For example, if a resolution in the 10 \mua  range as in \cite{lbi}
were possible, then from the second formula
of \eq{resl} one is in the $(M/M_\odot)\sim 10^{-7}$ regime, which would
be quite interesting. In \cite{lbi} earth-size baselines were used, so it
seems  that   for the
present purposes   achievement of such spectacular resolutions 
 would imply  extraterrestial arrangements\cite{lbi}.

 A positive discovery of primordial \bhs as compared
to merely setting upper limits  is of course the  more difficult 
 task, since one must
 eliminate other possible
angular averaging effects, both instrumental and natural. In this
connection it
should be noted that \eq{resl} is frequency
independent,  reflecting the achromatic behavior of light in \gvtl
fields, a feature that would not be expected for
many background effects.

We should stress that our limits, as in \eq{mlim}, are arrived at
by assuming that the \dkm is made of, or largely made of, \bhs
present at the formation of the CMB;
and so we can say nothing about the present existence of
 other objects like `stupendous' \bhsn,   perhaps
arising in the course of cosmic evolution \cite{stup}.

Finally we note, that although we speak about \bhsn, 
 the  arguments evidently  would apply to any massive,
 compact gravitating objects
making up the \dkm and present around the time of the formation of the CMB.
The objects might even be relatively extended, our estimates
should apply  as long as their effective \csss  doesn't become
comparable with that of \eq{css}.

\subsection{Power Spectrum vs. Angular Resolution}

The power spectrum, or equivalently directional  anisotropies,
  of the CMB has been a much discussed subject, with a
large literature \cite{hu}, To avoid possible confusion,
 we would like to stress that the resolution 
on the CMB,  which we have introduced  here, is not the same thing.
   We  contrast the two: the
power spectrum of the CMB  on the one hand, and the
 angular resolution in observing the CMB on the other.

That  the two cannot be the same is immediately evident when one
recognizes
that the two convey different sorts of information.
 The  power spectrum gives information on the nature of the
CMB itself, reflecting  primordial density fluctuations. On the
other hand,  the angular resolution 
 provides information on the `medium' between us and the CMB.
For the angular resolution question,  the role of
the CMB is simply to provide a `source' at high redshift. 

Of course, the existence of a finite resolution  affects analysis
of the
power spectrum. A finite resolution washes out structure as small
angles and leads to a softening of the
power spectrum at small angles or  high $l$. Indeed, once the
resolution is well understood
it can be deconvolved from the data to obtain the true power
spectrum.

The difference between the two is also made clear by considering
how they are determined.
To determine a power spectrum, one 
takes  the fourier transform of a  two-point correlation
function. The two-point correlation function, in 
turn, is an average over the data--in this case points or pixels
on  the sky.
The angular resolution on the other hand, concerns, in principle,
how well
 a single  point can be defined. No averaging is implied, and it
could
theoretically be determined from  one  point on the sky, if fine
enough
features existed.

\subsection{Open question}
In connection with this last remark, it is likely that are no
 very small angle features on the CMB, analogous to the
``acoustic peaks'', which could be used to demonstrate the desired
high angular resolution. A way around this problem--- to determine
a resolution when there are no suitable small targets--might be addressed by 
examining the  coherence of the radiation field over long
baselines. This
 is under study \cite{coh}.


\begin{thebibliography}{00}
\bibitem{zack} The principle of calculation we use is similar to that 
for other effects such as microlensing. See for example Eq 15 of 
E. Zackrisson and T. Riehm, Astronomy and Astrophysiscs {\bf 275}, 453, (2007),
arXiv:0709.1571  However,
it should be emphasized that there `high redshift'  is meant
a $z$ of 5 or 10, while in our discussion concerning the CMB we deal
with $z$ in the thousands. Thus the issues discussed in this reference
 are not  directly significant for our cosmological considerations, which
are dominated by the highest $z$.
\end{thebibliography}
\end{document}